\documentclass[aps,prl,reprint,superscriptaddress,showpacs,longbibliography]{revtex4-1}
\usepackage{graphicx,amsmath}
\usepackage{dcolumn}
\usepackage{bm}
\usepackage{color}

\begin{document}

\title{
The influence of measurement error on Maxwell's demon
}

\author{Vegard~S{\o}rdal}
\affiliation{Department of Physics, University of Oslo, 0316 Oslo, Norway}
%

\author{Y.~M.~Galperin}
\affiliation{Department of Physics, University of Oslo, 0316 Oslo, Norway}
\affiliation{A.~F.~Ioffe Physico-Technical Institute of Russian
Academy of Sciences, 194021 St. Petersburg, Russia}
\author{Joakim~Bergli}
\affiliation{Department of Physics, University of Oslo, 0316 Oslo, Norway}

\date{\today}

\begin{abstract}
In any general cycle of measurement, feedback and erasure, the
measurement will reduce the entropy of the system when information
about the state is obtained, while erasure, according to Landauer's
principle, is accompanied by a corresponding increase in entropy due
to the compression of logical and physical phase space. The total
process can in principle be fully reversible.
 A measurement error reduces the information obtained and the entropy decrease in the system.  The erasure still gives the same  increase in entropy and the total process is irreversible.  Another consequence of  measurement error is that a bad feedback is applied, which further increases the entropy production if the proper protocol adapted to the expected error rate is not applied. 
 We
 consider the effect of  measurement error on a realistic single-electron box 
  Szilard engine. We find the optimal protocol for the cycle as a function of the desired power $P$ and error $\epsilon$, as well as the existence of a maximal power $P^{\max}$.
\end{abstract}

\pacs{ 05.30.−d, 05.40.−a, 73.23.Hk, 74.78.Na}

\maketitle


Maxwell's demon was introduced as a thought experiment to illustrate
the statistical nature of the second law of thermodynamics
\cite{LeffRex2003}. The demon has very sharp powers of observation, so
it can detect the motion of individual molecules.
In addition, it can rapidly act on the basis of its observations and
thereby sort fast and slow molecules. This makes heat flow from the
cold to the hot side, apparently without the need for any work, in
contradiction to the second law of thermodynamics. For some
time it was thought that the act of
observation necessarily required some amount of work~\cite{szilard1929,brillouin}. The present
consensus~\cite{landauer,bennett1982} seems to be that the observation,
in principle, can be performed without work.
At the same time,
the erasure of the
information obtained, being a logically irreversible operation, also
is thermodynamically irreversible and has a necessary cost in terms of
work which is converted to heat. However, there is still some
controversy on this
point~\cite{Norton2011,Ladyman2013,PhysRevLett.102.250602}. 

Modern technology now enables us to be as accurate in observation and
quick in action as the imagined demon. Recently several
experiments which realize close analogies to the original thought experiment have been reported in a range of physical systems:
atoms~\cite{raizen2008,thorn2008,Raizen2009}, colloidal 
particles~\cite{toyabeNPhys2010,berut2012}, molecules~\cite{serreli2007},
electrons~\cite{Koski2014,Koski2015,chida2015}, and
photons~\cite{vidrighin2016}. This shift from
imagined to real experiments motivates us to study the impact of
measurement errors on the performance of experimental Maxwell's
demons. 

If there is some chance that the measurement result is wrong, it means
that the correlation between the state of the system and the
measurement device is not perfect. That is, the mutual information
between the two is less than the full information of the logical
states of the measurement device.  In~\cite{PhysRevLett.102.250602},
Sagawa and Ueda show that the traditional Landauer bound $ W\geq
T\ln2 $ (we use units where the Boltzmann constant $k_{\text{B}}=1$) only holds for a symmetric memory, and the total work
expended on measurement and memory erasure has a lower bound given by
the mutual information $I$ between the system and the measurement device,
\begin{equation}\label{modified_landauer}
W_{\text{measure}}+W_{\text{erase}}\geq TI.
\end{equation}
The r.h.s. is exactly the same as the heat which can be extracted from a
thermal bath using the information about the system.
 Although measurement errors
will give a reduced mutual information, we argue that it will not be possible to
reach equality in Eq.~(\ref{modified_landauer}) in this case.
To justify this,
consider the extreme case where the mutual
information $ I $ is zero, i.e., there is a $ 50\% $ chance that the
measurement is wrong. In this case the measurement can be done
reversibly without any work, but there will still be one bit of
information stored in the memory that has to be erased with a cost of
$T\ln2 $ according to Landauer.

To clearly show the difference between a true measurement error and
a process which saturates 
Eq.~(\ref{modified_landauer}), we will analyze a simple model.  
By distinguishing the degrees of freedom of a system into
information-bearing degrees of freedom (IBDF) and
non-information-bearing degrees of freedom (NIBDF)
\cite{Bennett2003501} the total entropy of the system can be separated
into two parts, the logical and the ``internal" entropy. Consider a
system  with a phase space $\mathcal P$. We divide the phase space in
subspaces $\mathcal P_i$, each of which corresponds to a specific
logical information stored. For a single bit, we have two subspaces,  which we
denote $ 0 $ and $ 1 $. With the probability distribution of the total
phase space denoted $ P(x) $, the  probability distribution of the logical states is
\begin{equation}\label{PL}
P_L(i)=\sum_{x\in \mathcal P_i}P(x),\quad i=0\vee 1
\end{equation}
and the conditional probability of the micro-state $ x $ given the logical state $ i $ is
\begin{equation}\label{Pcond}
P(x|i)=P(x)/P_L(i).
\end{equation}
The total entropy $ S $, logical entropy (information) $ H $ and conditional entropy $ S(\mathcal P_i|i) $ are  then given by
\begin{eqnarray}\label{entropy}
&&S = -\sum_{x}P(x)\ln P(x), \    
H=-\sum_{i}P_L(i)\ln P_L(i), \nonumber \\ &&
S(\mathcal  P_i|i) = -\sum_{x\in \mathcal{P}_i}P(x|i)\ln P(x|i).
\end{eqnarray}
The conditional entropy can be thought of as the internal physical
entropy of the distribution $ P(x|i) $ on $ \mathcal P_i $ for each of
the logical states $ i $. The average conditional entropy is $ S_{\text{in}}
= \sum_{i}P_L(i)S(\mathcal P_i|i) $, which we call the internal
entropy. It follows that we can write the total entropy as a sum
\begin{equation}\label{total_entropy}
S=H+S_{\text{in}},
\end{equation}
where $ H $ is associated with the IBDF, and $ S_{\text{in}} $ with the NIBDF.
\begin{figure}[t]
\centering
\includegraphics[width=0.8\linewidth]{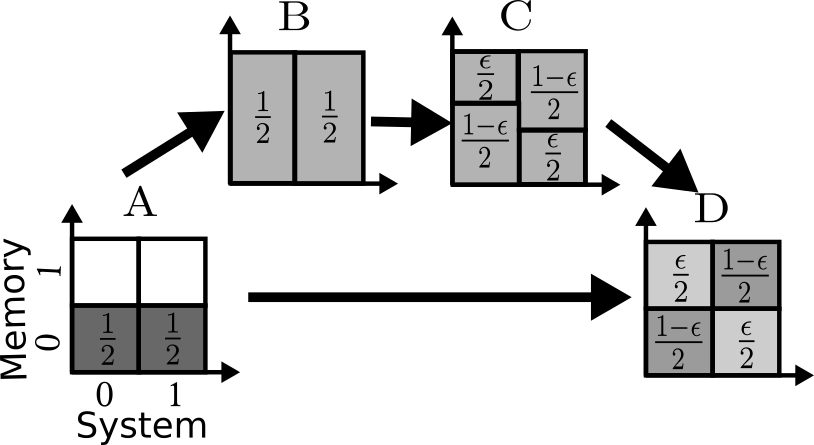}
\caption{A model system for analysis of the entropy flow.}
\label{fig:boxillustration}
\end{figure}

With this formalism we can analyze the model system shown in
Fig.~\ref{fig:boxillustration}. Assume that both the system and the
memory are represented by a standard Szilard engine, with a single
molecule in a box with a dividing wall which can be inserted, removed
and used as a piston. The phase space of each molecule is reduced to
one dimension by only considering the movement of the molecule in the
direction that the volume of the compartments expands/contracts and
ignoring the momentum, as all processes will be isothermal and
therefore the momentum distribution is constant. The relevant part of the
total phase space is then two-dimensional, and we represent the
position of the molecule in the system on the horizontal axis, and in
the memory on the vertical axis.  To calculate the entropy we use Eq.~(\ref{total_entropy}) and
the fact that the conditional entropy of a system uniformly
distributed in a given region of phase space is given by the logarithm
of the phase space volume. In Fig.~\ref{fig:boxillustration}A we then
have
%
\begin{equation}\label{1A}
H^A =   \ln 2, \   \
S^A_{\text{in}}=  -2\ln2,  \  \
S^{A}=-\ln2 .
\end{equation}

We perform a measurement on the system and store it in the memory. If
there is a  probability $ \epsilon $ that the measurement
gives the wrong result, we have a transition from
Fig.~\ref{fig:boxillustration}A to 1D. The total entropy of the state
shown in 1D is
\begin{equation}\label{1E}
H^D  =  \ln2 +S_\epsilon, \  \ 
S^D_{\text{in}} = -2\ln2, \  \
S^{D} = -\ln2+S_\epsilon,
\end{equation}
where $ S_\epsilon \equiv
-\epsilon\ln\epsilon-(1-\epsilon)\ln(1-\epsilon)$. The total entropy
in the transition from 1A to 1D is irreversibly increased by an amount
$ S_\epsilon $. Since the both the system and the memory have equal
probabilities of being in their two logical states, the logical
information in each is $H_{\text{System}}^D = H_{\text{Memory}}^D = \ln 2$. The mutual information between the system and memory is 
$$I^D =
H_{\text{System}}^D+H_{\text{Memory}}^D -H^D = \ln 2-S_\epsilon . $$

The state shown in 1D can also be reached reversibly while extracting
work if we consider the following steps (this process is also
considered in~\cite{sagawa2013}):
\begin{enumerate}
	\item [ A$ \to $ B] In the transition from 1A to 1B we
          isothermally expand the state 0 of the memory. This allows
          the particle to expand into the
          full volume of the memory. In this process work $W$ is
          performed by the system and heat $Q=W$ is taken from the
          reservoir. The entropy change is
	\[ \Delta S = W/T =\ln2 \] 
	with a corresponding entropy decrease in the reservoir. 
	\item[B $ \to $ C] We then perform a measurement on the
          system, and reinsert the partition wall in the memory
          according to the result obtained. There is no error in this
          measurement, and the correlation between the position of the
          dividing wall of the memory and the position (left/right) of
          the gas molecule of the system is perfect. Here $ \epsilon $
          is just a parameter that describes where we insert the
          divider in the memory. There is no entropy change.
	\item[C $ \to $ D] We then compress the divider of the memory
          isothermally back to the central position. In this process
          we have to perform work on the system, but an amount less than the work performed by it in the transition from
          1A to 1B. The entropy change is
	\[ \Delta S = W/T =  S_\epsilon - \ln2 .  \]
\end{enumerate}
In
our view, this process  does not represent a real measurement error,
which is irreversible and has an associated entropy production $ S_\epsilon $.
The final state of this process (1D) is the same as the one obtained
when there was a measurement error, but the whole process is
thermodynamically reversible, and the reduction of the environment
entropy is exactly the same as the increase of the system entropy. In
the process we have extracted net work from the thermal bath, so that
the work of measurement which enters Eq.~(\ref{modified_landauer}) is
$W_{\text{measure}} = -T S_\epsilon$ which is negative. Erasing the memory
requires $W_{\text{erase}} = T\ln 2$ according to the usual Landauer
principle, which gives 
$$W_{\text{measure}} +W_{\text{erase}} = T\ln 2-T S_\epsilon
=TI^D$$ which saturates the inequality~(\ref{modified_landauer}).

\begin{figure}[t]
\centering
\includegraphics[width=0.6\linewidth]{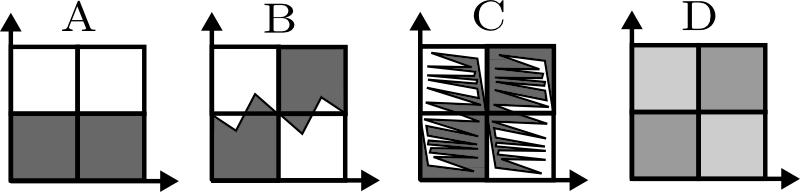}
\caption{How a system evolves from step A to D in Fig.\ref{fig:boxillustration} after a measurement error.}
\label{fig:irreversible}
\end{figure}

To get a deeper understanding of the irreversible nature of a
measurement with error, consider Fig.~\ref{fig:irreversible}.
In \ref{fig:irreversible}A we have the same initial state as before.
\ref{fig:irreversible}B shows the state just after the measurement was
performed. Most of the initial states in the phase space are mapped to
the correct final region, but a small fraction gets mapped to a
different region.  This corresponds to the cases where the result of
the measurement does not agree with the actual position of the system
molecule. If the system and the measurement device constitute an
isolated system during the operation, and no other degrees of freedom
are involved, the mapping from \ref{fig:irreversible}A to
\ref{fig:irreversible}B would be described by a deterministic
Hamiltonian evolution in time. Liouville's theorem then guarantees
that the entropy of the final state is the same as in the initial
state. If the evolution is affected by other microscopic degrees of
freedom in the device or the environment, which is certainly realistic
in most cases, the mapping would be stochastic, and depend on these
additional degrees of freedom. We can imagine that after B no further
changes of the logical states will occur. That is, the phase point
will never again cross the lines separating the different logical
states. In a short time the phase space region where the system can be
found will develop into some complicated shape
\ref{fig:irreversible}C, but for a closed system the entropy will
still be the same. Now we have to appeal to some coarse-graining
procedure. For a closed system the phase-space coarse-graining
introduced by Gibbs (see \cite{Ridderbos2002} for a recent
discussion). In the presence of some interaction with an environment,
coarse-graining over dynamical evolution
\cite{Blatt1959,lloyd1989}. In this way, the complex structure of the
accessible phase space in \ref{fig:irreversible}C is rendered
indistinguishable and replaced by the uniform distribution in
\ref{fig:irreversible}D. This step is irreversible and increases the
total entropy of the system by $S_\epsilon$ without any decrease in
entropy anywhere.


To see the effect of the entropy production in each measurement, we
will now analyze a model of an experimentally realized Szilard engine
\cite{Koski2014}. A single-electron-box (SEB) consisting of two
metallic islands connected by a tunnel junction. The existence of an an additional electron on one of the two islands can be
measured by the charge configuration of the box, and its state can
be controlled by gate voltages applied to the islands, giving
a time dependent potential difference $V(t)$ between the two
islands.  Work can be extracted from the system by the following
procedure
\begin{enumerate}
       \item Make the potential of the two islands equal, so
          that the probability of finding the extra electron is equal
          for the two islands.
	\item Perform a measurement, and if the extra electron is
          found on one island, quickly raise the potential of the
          other island to some value $ V_0 \equiv V(0^+)$.
	\item Move the potential of the island back towards
          zero according to some protocol $V(t)$.
\end{enumerate}
 There is a probability that the electron will tunnel to the other
 island, taking energy from thermal fluctuations. Whenever the
 electron occupy this island while the potential is decreasing, heat
 is extracted from the environment and converted to work. A model
 equivalent to this was previously analyzed~\cite{PhysRevE.88.062139}
 when there was no errors in the measurements, and the consequences of
 reduced mutual information (but with no entropy production associated
 with the measurement) were discussed~\cite{0295-5075-95-1-10005}. We
 imagine that we are continuously repeating the above steps, and we
 want to minimize the total entropy production rate when varying the
 driving protocol $V(t)$ and the time $ \tau $, at which we perform
 the next measurement and repeat the cycle.  In the limit
 $\tau\rightarrow\infty$, corresponding to quasistatic operation,
 the entropy production will vanish if
\begin{equation}\label{v0}
\left(e^{V_0/T} +1 \right)^{-1}=\epsilon .
\end{equation}
as shown in~\cite{0295-5075-95-1-10005}. This means that the
probabilities to find the electron on each of the islands are the same
as if there was thermal equilibrium at this value of $V_0$.

 While the entropy production rate can be zero when
 $\tau\rightarrow\infty$, we get a finite amount of work in an
 infinite time, which means that the power is
 zero. In~\cite{PhysRevE.88.062139} the problem of finding the $V(t)$
 and $\tau$ minimizing the entropy production rate with a given power
 $ P $ of heat taken from the reservoir was studied for the case
 $\epsilon=0$.  If there is an error in the measurement, the feedback
 operation $V(t)$ will have to be adapted to the expected error rate
 to minimize the entropy production rate. Extending the analysis to
 finite $\epsilon$ is principally not difficult, the details are
 described in the Supplementary information. It leads to an ordinary
 nonlinear differential equation which has to be solved
 numerically. We now present the main results of this analysis. The
 model has a parameter $\Gamma$ which determines the tunneling rate
 between the two islands, and we measure time in units of
 $\Gamma^{-1}$ and energy in units of temperature $T$.
\begin{figure}[t]
\centering
\includegraphics[width=1.1\linewidth]{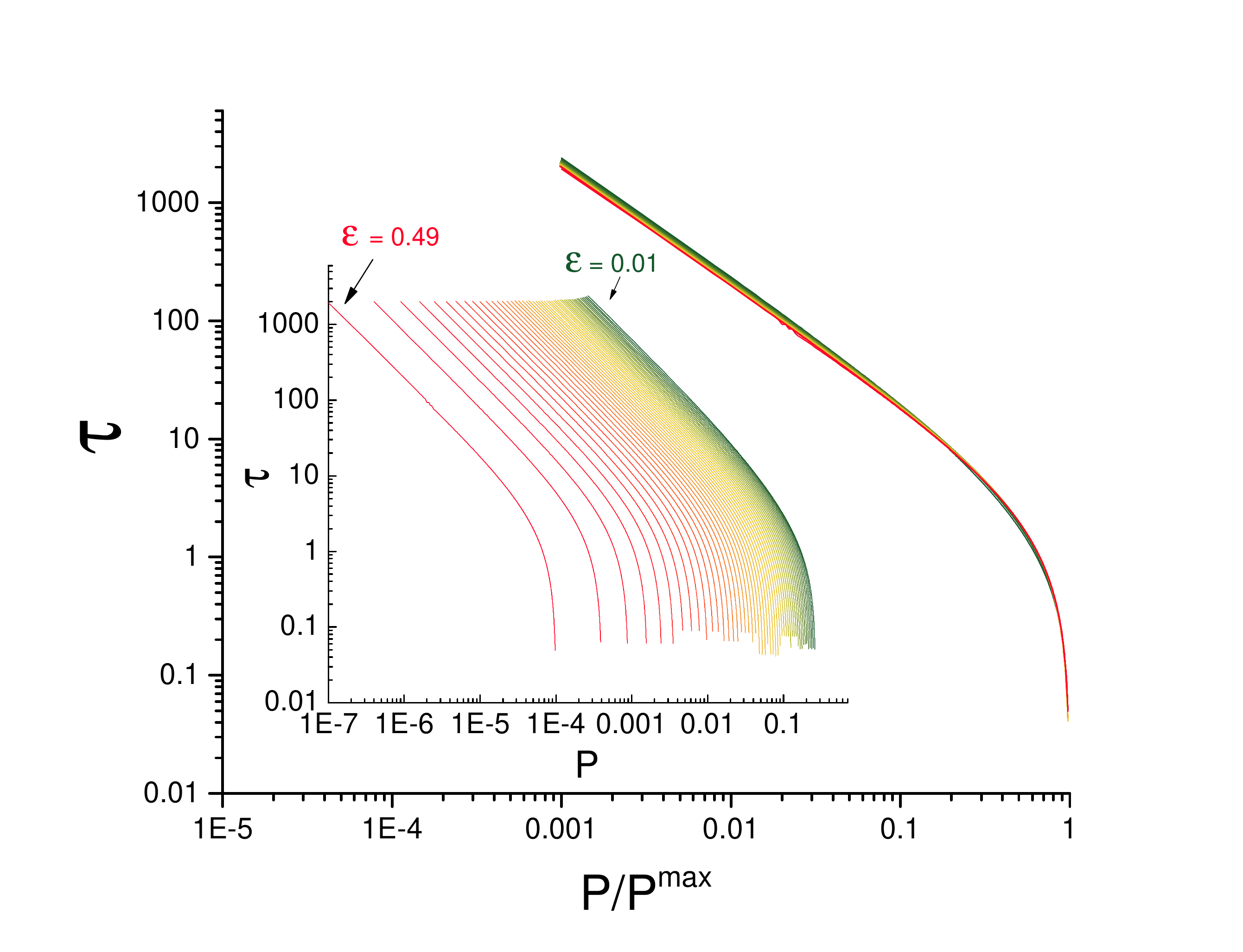}
\caption{The inset shows $\tau$ as a function of $P$ for different $\epsilon$. The main figure gives the scaled form of the same data, with $\tau$ as a function of $P/P^{\max}$.
\label{fig:entropytaup0}}
\end{figure}

In Fig.~\ref{fig:entropytaup0} (inset) we 
plot the optimal period
$ \tau $ as a function of the
power $P$ for selected values of the error $\epsilon $.
We find that there is a maximal
amount of power one can extract, $P ^{\max}(\epsilon)$, as $\tau$ approaches
0.  As $P$ approaches its maximum value $P ^{\max}$ the
period $\tau$ approaches
0 linearly:
$\tau\propto P^{\max}-P$.
  As one might expect, when the power $P$
goes towards zero, the optimal period $ \tau $ diverges to
infinity. In other words, when we approach reversibility by performing
the process in an infinite amount of time the power we can extract is
zero.  In the limit of low power $ P \to0 $ we find that $\tau = \left(\ln2-S_\epsilon\right)/P$, which we also confirm analytically in the supplement.
We have found two curious facts: 
(i) To a very good
approximation 
\begin{equation}\label{pmax}
  P ^{\max}(\epsilon)=\phi^{-1}(\epsilon -1/2)
  \sinh \left[\phi(\epsilon-1/2)\right],
\end{equation}
where $\phi=1.618$ is the golden ratio.  (ii) If $\tau$ is plotted as
a function of $P/P^{\max}$ the scaled graphs are close to collapsing
over the whole range of powers, as shown in
Fig.~\ref{fig:entropytaup0}.

\begin{figure}
\centering
\includegraphics[width=\linewidth]{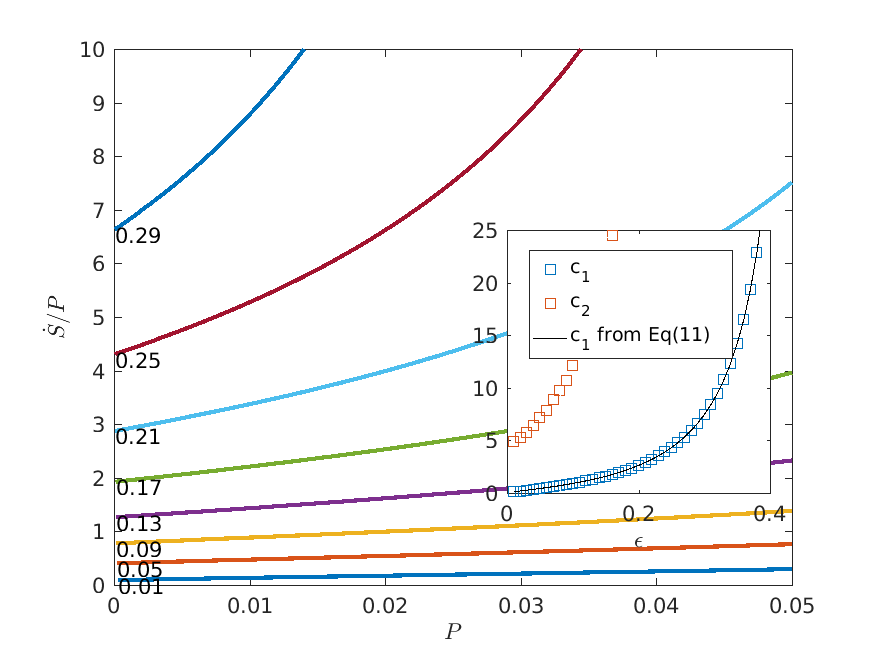}
\caption{$\dot S/P$ as a function of $P$ with labels on the curves giving $\epsilon$. For each curve, the value at $P=0$ and the slope of the tangent at that point will give  the coefficients $c_1$ and $c_2$ of Eq.~\eqref{sdot}. These are shown as functions of $\epsilon$ in the inset, together with $c_1$ from Eq.~\eqref{sdot2}.}
\label{fig:p0max}
\end{figure}

The entropy production rate diverges as 
$\dot{S}\propto \left(P^{\max}-  P \right)^{-1}$ when $P\rightarrow P^{\max}$, while it goes to zero for small $P$. In Ref. \cite{PhysRevE.88.062139} it was found that for $\epsilon=0$ and small $P$, $\dot S$ is proportional to $P^2$. We find that this is not true for finite $\epsilon$. We expand to second order, 
\begin{equation}\label{sdot}
\dot{S} =  c_1  P  + c_2  P ^2,
\end{equation}
where $c_1$ and $c_2$ are functions of $ \epsilon $. Plotting $\dot S/P$ as a function of $P$ (Fig. \ref{fig:p0max}) we get $c_1$ and $c_2$ as the intercept and slope of the tangent at $P=0$ (Fig.~\ref{fig:p0max}, inset).
In agreement with \cite{PhysRevE.88.062139} we find that $c_1$ goes to zero in the limit of $   P \to 0 $. The
entropy production rate $ \dot{S} $ is proportional to $   P ^2 $ for
error-free measurements, while it is proportional to $   P  $ if errors
are present. In fact we can predict $c_1$ by using the asymptotic result $\tau = (\ln2-S_\epsilon)/P$. According to Eq.~(10) of the supplement we have $\dot S = S_\tau/\tau-P$ where $S_\tau  = -p_\tau\ln p_\tau -(1-p_\tau)\ln(1-p_\tau)$ is the entropy at time $\tau$ with $p_\tau$ the probability to find the electron on one of the islands at time $\tau$. It is reasonable, and also confirmed by the numerical solution of the optimization problem (see Supplement), that at small $P$ and long time $\tau$ the potential will be brought  back to the initial value $V(\tau)=V_0$, so that final state will have equal probabilities for the electron to be found on either island, giving $S_\tau=\ln2$. We then get 
\begin{equation}\label{sdot2}
 \dot S = c_1 P \qquad\text{with} \qquad c_1 =S_\epsilon (\ln 2 - S_\epsilon)^{-1},
\end{equation}
which as shown in Fig.~\ref{fig:p0max} (inset) agrees perfectly with the numerical solution.

Let us summarize the main results: if we make an error in a
measurement, there is an associated net entropy production. This
applies to measurements of any type and with an arbitrary number of
outcomes. For a symmetric binary measurement where the probability of
error is $\epsilon$, the entropy increases by the amount $ S_\epsilon$.
This entropy
increase can be understood from a coarse-graining of either the phase
space (for a closed system) or the dynamical evolutions (for an open
system). We have investigated the consequences of a finite error
 on the optimal performance of a realistic Szilard engine
at finite (given) power. We found the existence of a maximal power
$P^{\max}$
which also exists
for error-free measurements, and which
decreases with increasing error. The entropy production
rate diverges as the maximal power is approached. For small power, the
entropy production rate is quadratic in $P$ in the absence of errors,
but becomes linear when errors are present. We also found the driving
protocol $V(t)$ and the time $\tau$ between measurements
that minimize the entropy production.
\acknowledgments
We are grateful to Jukka Pekola for illuminating discussions.

\newpage

\section{SUPPLEMENT}

\section{I. Details of the model and Calculations}

The model is the same as was studied previously 
\cite{PhysRevE.88.062139} without measurement errors. Here we briefly repeat the necessary definitions. 
Let $ p_1(t) $ and $ p_2(t) $ be the probabilities to find the system in state 1 (the right island) and 2 (the left island), respectively. The transitions between these two states are described by the rates $
\Gamma_{12} $ and $ \Gamma_{21} $, which satisfy detailed
balance $ \Gamma_{21}/\Gamma_{12}=e^{\Delta E/T} $  (note that since $\Delta E$ is a function of time, the rates will also be time dependent). The master equations are thus
\begin{eqnarray}\label{mastereq}
\dot{p}_1&=&-\Gamma_{12}p_1+\Gamma_{21}p_2=-\Gamma p_1 + \Gamma_{21},
\nonumber \\
\dot{p}_2&=&\phantom{-}\Gamma_{12}p_1-\Gamma_{21}p_2=-\Gamma p_2 + \Gamma_{12},
\end{eqnarray}
where $ \Gamma (t) \equiv \Gamma_{12}(t)+\Gamma_{21}(t) $. As in
\cite{PhysRevE.88.062139} we choose for simplicity $\Gamma$ to be
independent of time.  The energy of state $ i $ is denoted $ E_i(t) $,
and in the protocol described in the main text we have $ E_1(t)=0 $
and $ E_2(t)=V(t) $. The total work extracted during the period $ \tau
$ is
\begin{equation}\label{extracted_work}
W_{\text{ex}}=-\sum_{i=1}^{2}\int_{0}^{\tau}dt\, p_i\dot{E}_i,
\end{equation}
the change in internal energy of the system is
\begin{equation}\label{internal_energy}
\Delta U = \sum_{i=1}^{2}\left[p_i(\tau)E_i(\tau)-p_i(0)E_i(0)\right],
\end{equation}
and the transferred heat from the environment to the system is 
\begin{equation}\label{heat_transfer}
Q=\Delta U+W_{\text{ex}}=\sum_{i=1}^{2}\int_{0}^{\tau}dt\, \dot{p}_iE_i(t).
\end{equation}
The information
entropy associated with the measurement is $ H=-\sum_{i=1}^{2}p_i\ln
p_i $, and the entropy production is therefore $
\dot{H}=-\sum_{i=1}^{2}\dot{p}_i\ln p_i $. The change in information
entropy can be written as an integral
\begin{equation}\label{Shannon_entropy}
\Delta H =-\sum_{i=1}^{2}\int_{0}^{\tau}dt\, \dot{p}_i\ln p_i.
\end{equation}
Since $ p_1 = 1-p_2 $, we can relabel $ p_2\equiv p $, and write the entropy produced per cycle as
\begin{eqnarray}\label{SHannon_entropy2}
\frac{\Delta H}{\tau}&=&
-\frac{1}{\tau}\int_{0}^{\tau}dt~\dot{p}\ln\left(\frac{p}{1-p}\right).
\end{eqnarray}
The master equation \eqref{mastereq} can be expressed as 
\begin{equation}\label{pdot}
\dot p =  - p + \frac{1}{e^{V}+1}
\end{equation}
where from now on we will measure time in units of $\Gamma$ and energy in units of $T$. From this equation we can express 
\[
V = \ln\left(\frac{1}{p+\dot p}-1\right).
\]
The power is defined as the average heat extracted from the reservoir
per cycle $ \tau $, $ P=Q/\tau $, and can be written as
\begin{equation}\label{power}
P =\frac{1}{\tau}\int_{0}^{\tau}dt~\dot{p}V =\frac{1}{\tau}\int_{0}^{\tau}dt~\dot{p}~\ln\left(\frac{1}{p+\dot{p}}-1\right).
\end{equation}
We are interested in the optimal protocol for the measurement and erasure cycle. In this system the optimal protocol means finding the protocol $ V(t) $ and the  total time $\tau$ we should use on the cycle,
that minimize the entropy production rate given a measurement error $ \epsilon\in[0, \, 1] $ and a desired power $  P  $.
The total entropy production rate for perfect measurements is
\begin{equation}\label{total_entropy_production1}
\frac{\Delta S_{\text{tot}}}{\tau}=\frac{\Delta
	H}{\tau}- P .
\end{equation}
To study the effect of measurement errors, we have to add the entropy produced in the measurement, $ S_\epsilon = -\epsilon\ln\epsilon-(1-\epsilon)\ln(1-\epsilon)$, as discussed in the main text:
\begin{equation}\label{total_entropy_production2}
\frac{\Delta S_{\text{tot}}}{\tau}=\frac{\Delta
	H}{\tau}+\frac{S_{\epsilon}}{\tau}- P .
\end{equation}
We are interested in solutions where the power is given by a finite non-zero value, given by Eq.~(\ref{power}).

The initial condition is $ p(t=0) =\epsilon $. That is, there is a chance, $ \epsilon $, that the electron was on the island where  the potential  was  raised from $ V(0)=0 $ to $ V(0^+)=V_0 $, and thus preforming work on the system. We also set the value of the power, $  P  $, to see how the solutions depend on the power we want to extract.\\

Since $ S_\epsilon $ is a constant value that depends only on the initial condition,
it is sufficient to minimize the information entropy given in Eq.~(\ref{SHannon_entropy2}). Since we want to minimize it while keeping the power at a finite value $  P  $, we have to  introduce the Lagrange multiplier $ \lambda $ to obtain the functional
\begin{equation}\label{functional}
I=\frac{\Delta H}{\tau} + \lambda  P  = \frac{1}{\tau}\int_{0}^{\tau}dt~L(p,\dot{p},\lambda),
\end{equation}
with the Lagrangian
\begin{equation}\label{Lagrangian}
L(p,\dot{p},\lambda)=\left[-\ln\left(\frac{p}{1-p}\right)+\lambda\ln\left(\frac{1}{\dot{p}+p}-1\right)\right]\dot{p}.
\end{equation}
Using the Euler-Lagrange equation 
\begin{equation}\label{Euler-Lagrange}
\frac{\partial L}{\partial p}=\frac{\partial}{\partial t}\frac{\partial L}{\partial \dot{p}}
\end{equation}
we obtain the following second-order nonlinear ordinary differential equation:
\begin{equation}\label{differential_equation}
\ddot{p}=\frac{\dot{p}^2(\dot{p}+p-1/2)}{p(\dot{p}+p-1)+\dot{p}/2}\, .
\end{equation}
In order to solve this equation we need to impose a set of constraints to the solutions we want. The first constraint is that the power has to be a finite fixed value $  P  $, given by Eq.~(\ref{power}):
\begin{equation}\label{Power_constraint}
G(\tau,p,\dot{p})  \equiv  P  - \frac{1}{\tau}\int_{0}^{\tau}dt~\dot{p}\ln\left(\frac{1}{p+\dot{p}}-1\right) =0 .
\end{equation}
The second constraint comes from a consideration of the endpoint values of $ p(t) $. The initial condition of $ p(t) $ is given by $ p(0)=\epsilon $, but since the value of $ p(t) $ is not fixed at the endpoint $ p(\tau) $ we have a second constraint, $(\partial L/\partial \dot{p})_{t=\tau}=0 $, which can be written as
\begin{eqnarray}\label{endpoint_constraint}
F_1\left(\lambda,\tau,p\right)\equiv &&\lambda\left[\ln\left(\frac{1}{p+\dot{p}}-1\right)+\frac{\dot{p}}{(\dot{p}+p-1)(\dot{p}+p)}\right]
\nonumber \\ && 
-\ln\left(\frac{p_\tau}{1-p_\tau}\right)=0.
\end{eqnarray}
The third and final constraint is due to the fact that variation of Eq.~(\ref{total_entropy_production2}) with respect to the period $ \tau $ should be zero. It is given by
\begin{equation}
\frac{\partial\Delta S_\text{tot}}{\partial \tau}= \lambda\frac{\partial  P }{\partial\tau}-\frac{1}{\tau^2}\left(\Delta H  + S_\epsilon\right)+\frac{1}{\tau}\frac{\partial S_\tau}{\partial\tau}=0
\end{equation}
where 
\begin{eqnarray}
\frac{\partial  P }{\partial\tau} &=& \frac{\partial}{\partial\tau}\left[\frac{1}{\tau}\int_{0}^{\tau}dt~\dot{p}~\ln\left(\frac{1}{p+\dot{p}}-1\right)\right] \nonumber \\
&=& \frac{\dot{p}_\tau}{\tau}\ln\left(\frac{1}{p_\tau+\dot{p}_\tau}-1\right)-\frac{ P }{\tau}, 
\end{eqnarray}
and
\begin{eqnarray}\label{key1}
\frac{\partial}{\partial\tau}\frac{\Delta H}{\tau}&=&\frac{\partial}{\partial\tau}\left[-\frac{1}{\tau}\int_{0}^{\tau}dt~\dot{p}\ln\left(\frac{p}{1-p}\right)\right] \nonumber\\
&=&\frac{1}{\tau}\dot{p}_\tau\ln\left(\frac{1-p_\tau}{p_\tau}\right) -\frac{\Delta H}{\tau^2}.
\end{eqnarray}
The full equation for the third constraint is thus
\begin{eqnarray}\label{key2}
F_2(\lambda,\tau,p) &\equiv&
\left[\ln\left(\frac{1-p_\tau}{p_\tau}\right)+\lambda\dot{p}_\tau\ln\left(\frac{1}{p_\tau+\dot{p}_\tau}-1\right)\right] \nonumber \\
&-&\lambda  P -\frac{1}{\tau}\left[\Delta H + S_\epsilon\right]=0,
\end{eqnarray}
This constraint can be combined with the free-endpoint constraint by eliminating  the Lagrange multiplier $ \lambda $ to obtain
\begin{eqnarray}
&& F\left(\tau,p_\tau,\dot{p}_\tau\right)\equiv \ln\left(\frac{p_\tau}{1-p_\tau}\right)\left[ P (\dot{p}_\tau+p_\tau)(p_\tau+\dot{p}_\tau-1)
\right. \nonumber \\ && \left.
\qquad +\dot{p}^2_\tau\right]  
- \frac{S_\tau}{\tau}\left[\dot{p}_\tau+\ln\left(\frac{1}{p_\tau+\dot{p}_\tau}-1\right)
\right. \nonumber \\ && \left.  \qquad \qquad \times
(\dot{p}_\tau+p_\tau)(p_\tau+\dot{p}_\tau-1)\right]=0
\label{comb_const},
\end{eqnarray}
where $ S_\tau = \Delta H+ S_\epsilon = -p_\tau\ln p_\tau -(1-p_\tau)\ln(1-p_\tau)$ is the entropy of system at time $ t=\tau $.

It may seem surprising that the Lagrange multiplier $ \lambda $ disappears in the solution to the Euler-Lagrange equation, but this is because the entropy term in Eq.~(\ref{SHannon_entropy2}) is a complete integral. It is of course a state-function, that only depends of the initial and final values of $ p(t) $.
\begin{eqnarray}\label{key}
\Delta S&=&-\int_{0}^{\tau}dt~\dot{p}\ln\left(\frac{p}{1-p}\right) =-\int_{0}^{\tau}dp~\ln\left(\frac{p}{1-p}\right) \nonumber \\
&=&\bigg[-p\ln p - (1-p)\ln(1-p)\bigg]_0^\tau 
= S_\tau-S_0.
\end{eqnarray}
We use Euler's method to solve the second order differential equation in Eq.~(\ref{differential_equation}). Since it is a second order equation we have two constants that needs to be fixed ($\tau$ and $V_0$). We find these values as the roots of the two constraints in Eq.~(\ref{Power_constraint}) and Eq.~(\ref{comb_const}) by using Newton's method.

\section{II. Additional results}

Here we present some additional results on the optimal protocol. 

\subsection{The protocol $V(t)$}

\begin{figure}
	\centering
	\includegraphics[width=\linewidth]{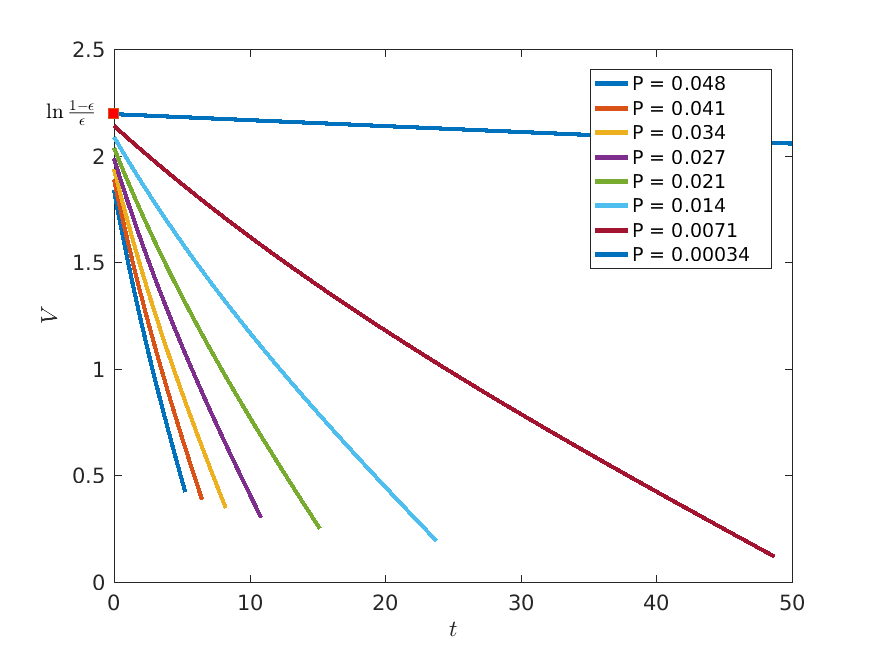}
	\caption{$V(t)$ for $\epsilon=0.1$ and several values of $P$.
		\label{fig:vt}}
\end{figure}

\begin{figure}
	\centering
	\includegraphics[width=\linewidth]{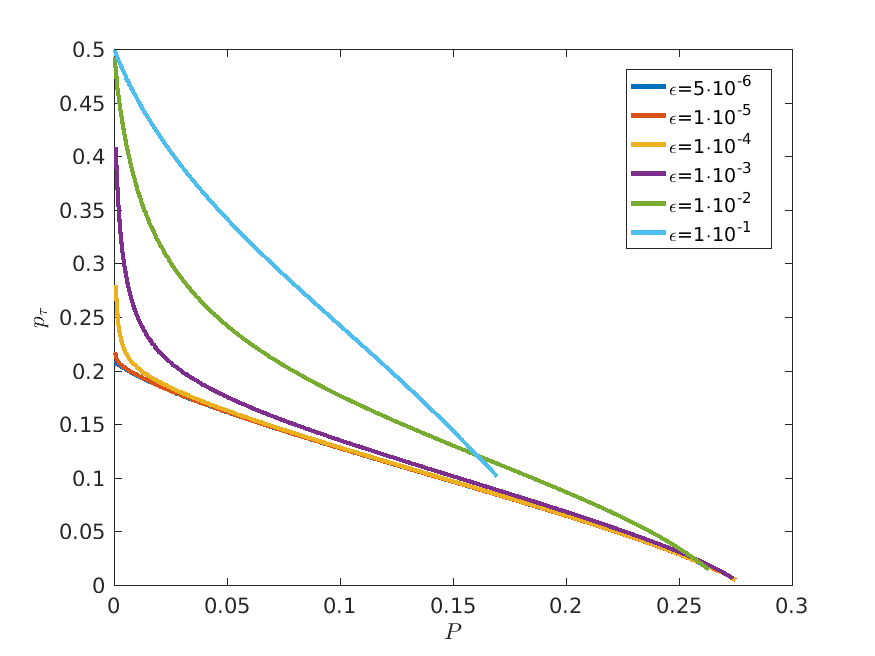}
	\caption{$p_\tau$ as function of $P$ for different $\epsilon$
		\label{fig:ptau}}
\end{figure}

The exact form of the function $V(t)$ which minimizes the entropy
production can only be found from the numerical solution of
Eq.~(\ref{differential_equation}). However, some limiting cases and
the dependence on the parameters $\epsilon$ and $P$ can be understood
to some extent. The numerical solution for $V(t)$ is shown in Fig.~\ref{fig:vt}
for $\epsilon=0.1$ and several values of $P$. We observe several
facts: i) The time $\tau$ before the protocol should be repeated
decreases with increasing $P$. ii) The initial value $V_0$ increases
with decreasing $P$. In the limit $P\rightarrow0$, we should have
according to Eq.~(8) of the main text $V_0 =
\ln\frac{1-\epsilon}{\epsilon}$, which is marked in the figure. We see
that the numerical results agree with this prediction. iii) The final value $V_\tau=V(\tau)$ depends on $P$ and goes to zero for small $P$. 
\begin{figure}
	\centering
	\includegraphics[width=\linewidth]{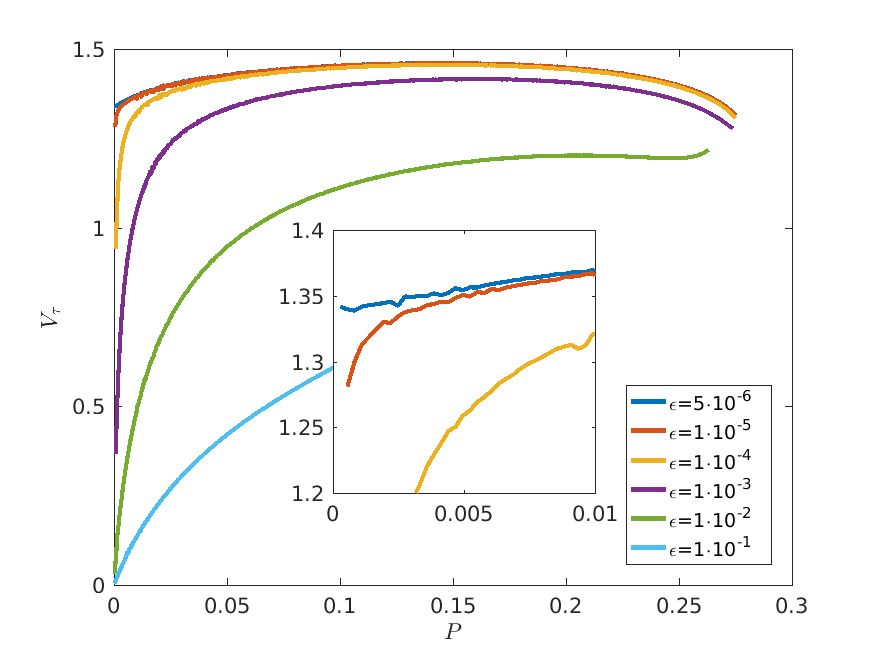}
	\caption{$V_\tau$ as function of $P$ for different $\epsilon$. The inset shows enlarged what happens for small errors and powers. 
		\label{fig:vtau}}
\end{figure}
Figure~\ref{fig:vtau} shows $V_\tau$ as function of $P$ for different
$\epsilon$. While it seems that for any finite $\epsilon$ we find
$V_\tau\rightarrow0$ as $P\rightarrow0$, we see that for small
$\epsilon$ one has to go to very small powers to see this, and for
most power $V_\tau$ is between 1 and 1.5. This indicates a singular
behaviour of the function $V_\tau(P,\epsilon)$ at $P=0$ and
$\epsilon=0$, and the limiting value will depend on how this point is
approached. In \cite{PhysRevE.88.062139} we found that $V_\tau=1.33$
for $\epsilon=0$ and small $P$. From Fig. \ref{fig:vtau} (inset) we
can see that this agrees well with what we would expect if we first
took the limit $\epsilon\rightarrow0$ and then $P\rightarrow0$. The
same singularity is reflected in the probability $p_\tau$ to find the
electron on the opposite island at time $\tau$ from the one it was
measured at time 0 as shown in Fig.  \ref{fig:ptau}. For all finite
$\epsilon$ we have $\lim_{P\rightarrow0}p_\tau = 0.5$, but for small
$\epsilon$ this only happens at very small $P$.

\subsection{The maximal power, $P ^{\max} $}

We can always find the value of $P ^{\max} $ from the numerical solution of Eqs.~\eqref{differential_equation}, \eqref{Power_constraint}, \eqref{comb_const} by determining when $\tau$ becomes 0. But we can also derive a single transcendental equation which determines $P ^{\max} $ , and in the case of error-free measurements
we can also solve it analytically.

By taking the limit as $ \tau\to0 $
in 
Eq.~(\ref{power}) we find 
that
\begin{equation}\label{p0}
P ^{\max}=  V_0\dot p_0 = V_0\left(\frac{1}{e^{V_0}+1}-\epsilon\right).
\end{equation}
which expresses $ P ^{\max}$ in terms of $V_0$. Consider Eq. \eqref{comb_const} when $\tau\rightarrow0$. Since the other terms are finite, the only way to avoid a divergence of the last term is that expression in brackets is zero. For $\tau=0$ we have  $p_\tau=\epsilon$ and with Eq. \eqref{pdot} we find that 
$V_0$ satisfies the equation
\begin{equation}\label{v0e}
1+(1-V_0)e^{V_0}-\epsilon(e^{V_0}+1)^2=0.
\end{equation}
For $ \epsilon=0 $ we find that the maximum power is given by the
Lambert W function
\begin{equation}\label{lambert}
P ^{\max}=W(e^{-1})= 0.27846\dots
\end{equation}
with the initial value of the potential $ V_0=1+W(e^{-1})$. This
analytical result is in perfect agreement with our numerical one.

Curiously, a good
approximation to this plot is given by
\begin{equation}\label{pmax}
P ^{\max}=\frac{\epsilon-1/2}{\phi}\sinh \left(\phi(\epsilon-1/2)\right).
\end{equation}
The difference between the true and approximate solution is only $ 10^{-4} $ for $\epsilon=0$:
\begin{equation}\label{pmaxApprox}
P ^{\max}=\frac{1}{2\phi}\sinh \left(\frac{\phi}{2}\right)=0.27817\dots
\end{equation}

\subsection{The dependence of $\tau$ on $P$ for small $P$}

When $P=0$ we can assume the system to always be in equilibrium at the given value of $V$, which means that $p=p_a = \left(e^V+1\right)^{-1}$.
We assume for small $P$ we have $p=p_a +\mathcal{O}(P) $, and  that $\tau=A/P$. Inserting into Eq. \eqref{power} and expanding in $P$ we find that it becomes

\[
1 = \frac{1}{A}\int_0^\infty dt V\dot p_a + \mathcal{O}(P)
\]
Using the fact found above that $V(\infty)= \lim_{P\rightarrow0}V_\tau = 0$ (at leat for finite $\epsilon$) and that $ \lim_{P\rightarrow0}V_0 =
\ln\frac{1-\epsilon}{\epsilon}$, we get 

\[
A = \int_0^\infty dt V\dot p_a = -\int_0^{V_0} dV V\frac{dp_a}{dV} = \ln2-S_\epsilon
\]

\end{document}